\documentclass[pss,fleqn,embeddedheads]{w-art}
\usepackage{times}
\usepackage{w-thm}
\usepackage[]{graphicx}
\begin{document}
\DOIsuffix{theDOIsuffix}
\Volume{XX}
\Issue{1}
\Copyrightissue{01}
\Month{01}
\Year{2004}
\pagespan{1}{}
\Receiveddate{\sf zzz} \Reviseddate{\sf zzz} \Accepteddate{\sf
zzz} \Dateposted{\sf zzz}

\title[]
{Superconductivity in the two-dimensional Hubbard model?}

\author[D. Baeriswyl]{D. Baeriswyl\footnote{Corresponding
     author: e-mail: {\sf dionys.baeriswyl@unifr.ch}, 
Phone: +41\,26\,300\,9141, Fax: +41\,26\,300\,9758}}
\address{Department of Physics, University of Fribourg, 
Chemin du Mus\'ee 3, CH-1700 Fribourg, Switzerland}
\author[D. Eichenberger]{D. Eichenberger}
\author[B. Gut]{B. Gut}
\begin{abstract}
A refined variational wave function for the two-dimensional repulsive
Hubbard model is studied numerically, with the aim of approaching the 
difficult crossover regime of intermediate values of $U$. The issue
of a superconducting ground state with $d$-wave symmetry is 
investigated for an average electron density $n\approx 0.8$ and for
$U=8t$. Due to finite-size effects a clear-cut answer to this fundamental 
question has not yet been reached.
\end{abstract}
\maketitle                   


\renewcommand{\leftmark}
{D. Baeriswyl, D. Eichenberger and B. Gut: 
Superconductivity in the two-dimensional Hubbard model}

\section{Introduction}

Strong correlations between electrons play a fundamental role
in the cuprates. In fact, there is ample evidence for a crossover from a 
doped Mott insulator to a correlated Fermi liquid, as holes are introduced
into the CuO$_2$ planes.
Two limiting regimes can be described in simple terms, an antiferromagnetic
Mott insulator at half filling ($x=0$) and a rather conventional Fermi 
liquid at high doping levels ($x>0.25$). Many difficult problems remain
to be solved, especially concerning the nature of the underdoped region
(the so-called pseudogap phase), the normal phase around optimal doping
(``marginal Fermi liquid'') and the ``physical mechanism'' leading to
the superconducting phase for $0.05<x<0.25$. In this note we discuss two
specific issues: on the one hand, the nature of the crossover between the
underdoped and the overdoped regimes, on the other hand, the possibility 
of superconductivity originating from purely repulsive interactions.
We use variational wave functions for the one-band Hubbard model in two
dimensions, keeping in mind that both the model and the method do provide  
insight, but at the same time are not sufficient for developing
a complete theory of the layered cuprates.

The Hubbard Hamiltonian $\widehat H=-t\widehat T+U\widehat D$ is composed 
of two terms with conflicting tendencies, the hopping term
\begin{equation}
\label{hopping}
\widehat T=\sum_{\langle i,j\rangle ,\sigma}
(c^\dag_{i\sigma}c_{j\sigma}+c^\dag_{j\sigma}c_{i\sigma})\; ,
\end{equation}
where $c^\dag_{i\sigma}$ creates an electron at site $i$ with spin $\sigma$,
and the number of doubly occupied sites
\begin{equation}
\label{double}
\widehat D=\sum_i n_{i\uparrow}n_{i\downarrow}\; ,
\end{equation}
where
$n_{i\sigma}=c^\dag_{i\sigma}c_{i\sigma}$. The notation 
$\langle i,j\rangle$ means
that hopping is restricted to neighboring sites. 

If the on-site repulsion $U$ vanishes, the electrons occupy delocalized
Bloch orbitals and set up a filled Fermi sea in the ground state
$|\psi_0\rangle$, given by
\begin{equation}
\label{bloch}
|\psi_0\rangle=
\prod_{{\bf k},\sigma,\varepsilon_{\bf k}<\varepsilon_F}
c_{{\bf k},\sigma}^\dag|0\rangle\; .
\end{equation} 
In contrast, for a vanishing hopping amplitude $t$ 
(and $U>0$, or for finite $t$ and $U\rightarrow\infty$ ), the ground
state $|\psi_{\infty}\rangle$ for $N=2M$ electrons is an arbitrary 
superposition of
configurations where the electrons are localized on lattice sites without
any double occupancy,
\begin{equation}
\label{local}
|\psi_{\infty}\rangle=
\sum_{\begin{array}{l}i_1,\cdots,i_M\\j_1,\cdots,j_M\end{array}}
\varphi(i_1,\cdots,i_M;j_1,\cdots,j_M)\;
c_{i_1,\uparrow}^\dag\cdots c_{i_M,\uparrow}^\dag
c_{j_1,\downarrow}^\dag\cdots c_{j_M,\downarrow}^\dag|0\rangle\; ,
i_k\neq j_l\; .
\end{equation} 
For finite values of $U$ and $t$ the delocalizing tendency of $t$ and
the localizing tendency of $U$ compete. One expects a crossover to occur around
some critical value $(U/t)_c$ between a ground state linked to $|\psi_0\rangle$
for $U<U_c$ and one linked to $|\psi_\infty\rangle$ for $U>U_c$. This picture
is captured by two types of variational wave functions, the Gutzwiller
ansatz \cite{gutzwiller}
\begin{equation}
\label{gutz}
|\psi\rangle_G=e^{-g\widehat D}|\psi_0\rangle
\end{equation}
and its counterpart \cite{baeriswyl}
\begin{equation}
\label{db}
|\psi\rangle_B=e^{-h\widehat T}|\psi_{\infty}\rangle\; ,
\end{equation}
where $g$ and $h$ are variational parameters. Depending on the relative values
of the minima of
\begin{equation}
\label{energy}
E_G(g)=\frac{\langle\psi_G|\widehat H|\psi_G\rangle}
{\langle\psi_G|\psi_G\rangle}\; ,\; \;
E_B(h)=\frac{\langle\psi_B|\widehat H|\psi_B\rangle}
{\langle\psi_B|\psi_B\rangle}\; ,
\end{equation}
one or the other of the two ground states is favored.
\section{Mott transition and localization}
For the case of a half-filled band, $n=N/L=1$ 
($L$ is the total number of sites), the two variational states defined above
can be distinguished by their sensitivity with respect to changes in boundary
conditions. Using an early argument of Kohn \cite{kohn}, it can be readily 
shown
that $|\psi_G\rangle$ is metallic (the Drude weight is finite) and 
$|\psi_B\rangle$ is insulating (vanishing Drude weight) \cite{dzierzawa}. 
Thus the 
delocalization-localization crossover manifests itself in this approach as a
sharp (Mott) metal-insulator transition at a critical value $U_c$, which
turns out to be of the order of the bandwidth.
The distinction between the two regimes
is blurred if we allow for antiferromagnetic ordering, which is expected to
set in at arbitrarily small $U$ for the square lattice. In this case, a 
possible signature would
be the size of the local moment, which is small in the ``delocalized''
spin-density-wave regime at small $U$, increases strongly in the crossover
regime and saturates for $U\rightarrow\infty$.  

The insulating cuprates can be well understood in terms of the 
antiferromagnetic
Heisenberg model, the strong coupling limit of the half-filled Hubbard
model, provided that ring exchange around the plaquettes is also included.
The presence of this term indicates that $U$ is not much larger than the
bandwidth. In fact, the analysis of the spin waves observed in neutron 
scattering 
experiments yields a value of $U$ of the order of $12t$ (twice the bandwidth)
\cite{aeppli}. 
According to our variational analysis this puts these materials inside 
the localized regime, but not so far that double occupancy can be simply
discarded. At half filling, cuprates can thus be considered as 
``weakly localized'' Mott insulators. The good agreement between theoretical
predictions made on the basis of the (two-dimensional) Hubbard model and
the observed magnetic exitation spectrum in the layered cuprates does not
imply that long-range Coulomb interactions are 
absent in these materials; these are just not seen at half filling. 
The addition of non-local Coulomb terms would essentially lead to 
slightly different exchange constants. 

An interesting and largely unexplored issue is the location of the crossover
regime (as a function of $U$) away from half filling. The variational
procedure should also be useful in this limit, although the presence of
holes poses an additional challenge. The distinction between the localized
and delocalized regimes is less clear-cut for $n<1$ than it is at
half filling because the Drude weight is expected to remain finite due to the 
motion of holes. A further complication
arises from the fact that the Hubbard Hamiltonian represents an effective 
model, in which many degrees of freedom are subsumed in some average way by 
the parameters. Quite generally,
we expect that screening will become more
and more efficient as doping is increased and thus will lead to a reduced 
value of $U$. The appearance of a conventional
Fermi liquid above a hole concentration of about 25\% agrees with such 
a picture.

\section{$d$-wave superconductivity}

The celebrated BCS theory of superconductivity is based on an effective 
attraction between electrons, caused by phonon exchange. Therefore it is
not surprising that the application of the BCS mean-field approximation to
the repulsive Hubbard model does not yield any energy gain due to a pairing
instability. To see this, we decompose the on-site term,
\begin{equation}
\langle n_{i\uparrow}n_{i\downarrow}\rangle \approx
\langle n_{i\uparrow}\rangle\langle n_{i\downarrow}\rangle
+\langle c_{i\uparrow}^\dag c_{i\downarrow}\rangle
\langle c_{i\uparrow}c_{i\downarrow}^\dag\rangle
+\langle c_{i\uparrow}^\dag c_{i\downarrow}^\dag\rangle
\langle c_{i\downarrow}c_{i\uparrow}\rangle\; ,
\end{equation}
and notice that in the absence of antiferromagnetism the first term is just
$\frac{1}{4}n^2$, while the second term vanishes. The third term, responsible
for a superconducting instability in BCS theory, is nonnegative,
positive for an order parameter with $s$-wave symmetry and zero for
$d$-wave symmetry. We conclude that in the BCS approximation
superconductivity with
$d$-wave symmetry (or with other similar order parameters) is neither 
promoted nor suppressed by the on-site repulsion. Therefore we have to go
beyond the mean-field level to find out whether a superconducting
ground state is conceivable in the repulsive Hubbard model. This is a
very difficult problem since it necessitates a reliable calculation
of the correlation energy. 

The question of superconductivity in the repulsive Hubbard model 
(and the related $t$-$J$ model) has
been attacked by several different methods, the perturbative Renormalization
Group (RG) \cite{zanchi, halboth}, numerical calculations of the renormalized 
scattering vertex \cite{scalapino} and variational wave functions
\cite{gros,shiba,giamarchi,paramekanti,sorella}.  
All these approaches are consistent with $d$-wave superconductivity in some
doping range, but most of them have a very limited range of applicability.
Thus the perturbative RG methods \cite{zanchi, halboth}
are only valid for very small values of $U$
and cannot provide a quantitative answer for the energy gap or the
condensation energy. Conversely, studies of the $t$-$J$ model 
\cite{gros, shiba, paramekanti, sorella} can only be applied to the large
$U$ limit of the Hubbard model. Unfortunately, reliable results for the 
crossover region of the two-dimensional Hubbard model are still scarce.

\section{A refined variational wave function}

In our own work \cite{eichenberger} we use a variational ansatz, which
is a combination of Eqs.~(\ref{gutz}) and (\ref{db}) and constructed in such
a way as to approach the crossover region from the Fermi liquid side and 
to allow for a superconducting ground state with $d$-wave symmetry,

\begin{equation}
\label{wf_def}
\qquad\qquad\vert\Phi\rangle=e
^{-h\hat T}e^{-g\hat D}\vert d\mbox{BCS}\rangle\ .
\end{equation}
Here the double occupancy is first partially suppressed 
(variational parameter g), and subsequently both the hopping of holes and the 
kinetic exchange are enhanced (variational parameter h). The additional
parameter $h$ 
reduces significantly the total energy, in agreement with similar earlier
work \cite{otsuka, michela}. The parent state $\vert d\mbox{BCS}\rangle$ is a 
BCS state with $d$-wave symmetry, {\it i.e.,}  
\begin{eqnarray}
\qquad u_{\bf{k}}^2=
\frac{1}{2}(1+\frac{\epsilon_{\bf{k}}-\mu}{E_{\bf{k}}}),
\qquad u_{\bf{k}}v_{\bf{k}}=\frac{\Delta_{\bf{k}}}{2E_{\bf{k}}}\ ,\nonumber
\end{eqnarray}
with
\begin{eqnarray}
E_{\bf{k}}&=&\sqrt{(\epsilon_{\bf{k}}-\mu)^2+\Delta_{\bf{k}}^2},
\qquad\Delta_{\bf{k}}=\Delta\cdot(\cos{k_x}-\cos{k_y}).\nonumber
\end{eqnarray}
Two further variational parameters are included in $\vert d\mbox{BCS}\rangle$,
the BCS gap $\Delta$ and a ``chemical potential'' $\mu$ 
that allows to fix the average number of electrons. We emphasize that $\mu$ 
is not identical to the true chemical potential. 
\begin{figure}[h]
\vspace{1.5cm}
\begin{center}
\includegraphics[width=10cm]{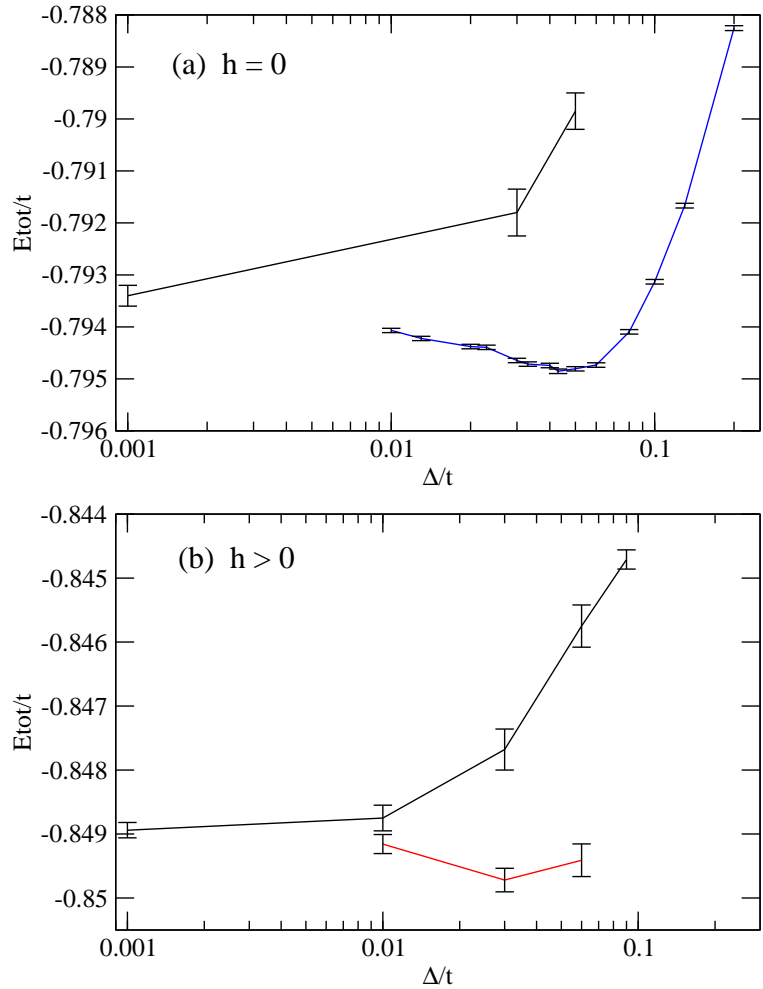}
\vspace{1cm}\caption{\label{Tot_Energy} $(a)$: Total energy per site of the 
Gutzwiller-type wave function. $(b)$: Total energy per site of the refined 
variational wave function. In both figures the lower curve corresponds to a 
precisely fixed number of electrons, while the upper curve represents the 
``grand canonical'' case with $\mu$ tuned to represent a fixed
average density. The calculation was carried out on a 8x8 square lattice for 
$n=0.8125$ and $U=8t$.}
\end{center}
\end{figure}

The variational parameters are determined by minimizing the total energy 
(expectation value of the Hamiltonian). The natural way to get this energy 
for a given density is to project the parent state on a fixed number of 
electrons, working either in real space or in momentum space, and then to 
compute the energy using a Monte Carlo simulation. For the case 
$h=0$ (Gutzwiller type wave function), the parent state is easily written as a 
superposition of configurations in real space, where the number of doubly 
occupied sites $\hat D$ is diagonal. The total energy, given by the lower 
curve in Fig. \ref{Tot_Energy}$(a)$, has a pronounced minimum around $0.05t$. 
The condensation energy is of the order of $0.001t$.

The case $h>0$ is more demanding because the operators $\hat T$ and $\hat D$ 
are diagonal with respect to different bases. In a first approach, we have
written the BCS state as 
a superposition of momentum space configurations with a fixed number
of electrons. A discrete Hubbard-Stratonovich transformation is used to 
decouple the fermion operators in the Gutzwiller projector. The price to pay 
is a summation over ``Ising spin'' configurations, in addition to the summation
over the momentum space configurations. This leads to a weight which is no 
longer positive definite in the Monte Carlo simulation. The resulting
sign problem becomes severe as the gap becomes large. The result is 
represented by the lower curve of Fig. \ref{Tot_Energy}$(b)$. In the range 
where our results are valid (do not suffer sign problems),  
the energy first decreases, reaches a 
minimum at about $0.03t$ and then increases.  For $\Delta=0.06t$ 
the average value of the sign is about $0.3$. Results for larger values of
$\Delta$ cannot be trusted. 

An alternative approach is to start with an unprojected parent state (i.e., 
with a fixed ``chemical potential'' instead of a fixed number of electrons).
In this case only the summation over ``Ising spin'' configurations has to be 
performed, but the ``chemical potential'' has to be fixed to get the right 
average density. Interestingly, working with this wave function 
allows to escape the minus sign problem. The resulting total energy is 
represented by the upper curve of Fig. \ref{Tot_Energy}$(b)$. The qualitative 
behaviour of the energy is quite different for the unprojected wave function; 
the minimum energy is located at $\Delta=0$! This result is confirmed 
by the same calculation carried out for $h=0$ (see Fig. \ref{Tot_Energy}$(a)$, 
upper curve). 

The two approaches should be equivalent in the thermodynamic limit. 
Unfortunately, for the system size studied so far ($8\times 8$), 
finite size effects seem to be rather large and, especially, lead
to two conflicitng results. Larger system sizes together with finite size 
scaling will be required before we can reach any definite conclusion 
about the fundamental issue of superconductivity in the repulsive Hubbard 
model.

\begin{acknowledgement}
This work was supported by the Swiss National Foundation
through the National Center of Competence in Research ``Materials with
Novel Electronic Properties-MaNEP'' and through grant no. 200020-105446. 
\end{acknowledgement}

\end{document}